\documentclass[prl,showpacs,floatfix,twocolumn]{revtex4}

\pdfoutput=1

\usepackage{graphicx}

\newcommand{\STNO}{SrTi$_{0.875}$Nb$_{0.125}$O$_3$}
\newcommand{\BTNO}{BaTi$_{0.875}$Nb$_{0.125}$O$_3$}

\begin{document}

\title{Local structural origins of the distinct electronic 
properties of Nb-substituted SrTiO$_3$ and BaTiO$_3$}

\author{
Katharine Page,$^1$
Taras Kolodiazhnyi,$^2$
Thomas Proffen,$^3$
Anthony K. Cheetham,$^4$ and 
Ram Seshadri$^1$
}
\affiliation{
$^1$Materials Department, University of California, 
Santa Barbara,  CA 93106, USA\\
$^2$National Institute for Materials Science,  1-1
Namiki, Tsukuba, Ibaraki, 305-0044 Japan\\
$^3$Los Alamos National Laboratory, Lujan Neutron Scattering Center,
MS H805, Los Alamos, NM 87545, USA\\
$^4$Department of Materials Science and Metallurgy, University of
Cambridge, CB2 3QZ, UK
}

\begin{abstract}
Near or less than 10\% Nb substitution on the Ti site in perovskite SrTiO$_3$ 
results in metallic behavior, in contrast to what is seen in BaTiO$_3$. 
Given the nearly identical structure and electron
counts of the two materials, the distinct ground states for low substitution
have been a long-standing puzzle.  Here we find from neutron studies of average
and local structure, the subtle  yet critical difference that we believe
underpins the distinct electronic  properties in these fascinating materials.
While \STNO\/ possesses a distorted non-cubic structure at 15\,K, the BO$_6$ 
octahedra in the structure are regular. \BTNO\/ on the other hand shows 
evidence for local cation off-centering whilst retaining a cubic structure.
\end{abstract}

\maketitle

\begin{figure}[h]
\centering\includegraphics[width=6cm]{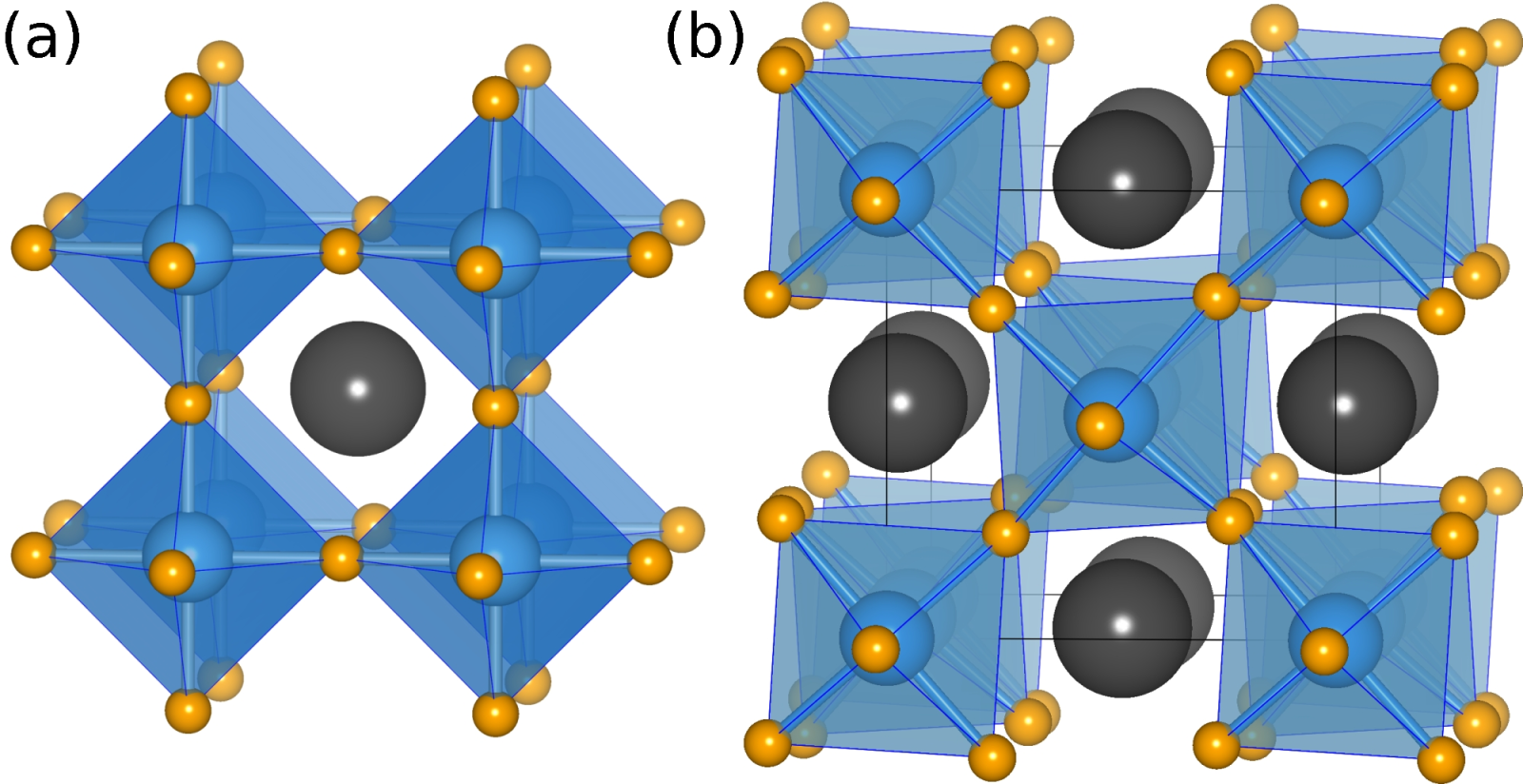}\\ 
\caption{(Color online.) Crystal structures of SrTiO$_3$ at (a) 
room temperature when it is cubic, and (b) at low temperatures when it is 
tetragonal. The structural parameters used to depict (b) were obtained from 
Rietveld refinement of the 15\,K neutron diffraction data of \STNO.
\label{fig:structures}} 
\end{figure}

Small substitutions of Nb for Ti in the perovskite SrTiO$_3$ give rise to
electronic  conductivity,\cite{Nb_STO_Plamsons,Nb-STO_ThinFilms} and even 
superconductivity.\cite{Nb_SrTiO3_supercond} Nb-substituted SrTiO$_3$ is 
widely used as a conducting substrate for epitaxial perovskite thin films.
BaTiO$_3$ which has a  ferroelectric ground state,\cite{Megaw} also substitutes
Nb on the Ti site, and is known to form almost a complete solid solution with
BaNbO$_3$. Despite possessing the same cubic perovskite structure, depicted 
in Fig.\ref{fig:structures}(a), (at least for greater than 10\,atom-\% 
substitution) and identical electron counts, the Sr compounds display metallic 
behavior while the Ba compounds are localized up to at least 20\% 
substitution.\cite{Marucco_solidsol} 

In order to unravel the possible structural underpinnings of the distinct 
behavior of the end-members in this solid solution series, we have prepared
sample compositions with 12.5\,\% Nb-substitution: 
SrTi$_{0.875}$Nb$_{0.125}$O$_3$ (STNO) and 
BaTi$_{0.875}$Nb$_{0.125}$O$_3$ (BTNO), and studied their average and local
structures using time of flight neutron diffraction measurements. 

Samples were prepared from high purity (99.99\%) 
Ba(Sr)CO$_3$, TiO$_2$ and Nb$_2$O$_5$ powders purchased from Sigma Aldrich. 
Pellets of stoichiometric amounts of the starting materials were
calcined at 1100\,$^\circ$C for 20\,h in flowing H$_2$, at a flow rate of 
50\,cm$^3$/min, with a second regrinding/repelletization at 1350\,$^\circ$C 
for 20\,h in flowing H$_2$. Four-probe resistivity was measured on 95\% dense 
ceramic samples cut into $2\times2\times9$ mm$^3$ parallelepipeds.  Magnetic
susceptibility in the 2\,K to 400\,K range was measured with a QD MPMS XL
magnetometer. High-$Q$ resolution neutron powder diffraction data were 
collected on the \STNO\/ and \BTNO\/
samples on the NPDF instrument at the Los Alamos National Laboratory Lujan
Neutron Scattering Center\cite{proffen_npdf} at room temperature and at 15\,K. 
Rietveld refinement of the diffraction data was carried
out in the \textsc{gsas-expgui}\cite{gsas-expgui} suite of programs.  The 
experimental pair distribution function was extracted from neutron total 
scattering data using the program \textsc{pdfgetn}.\cite{PDFGETN} 
The wavevector cutoff $Q_{max}$ used for the transform was
40\,\AA$^{-1}$\/.  PDF refinements were carried out using the \textsc{pdfgui} 
program.\cite{PDFgui} 

\begin{figure}[t]
\centering\includegraphics[width=6.5cm]{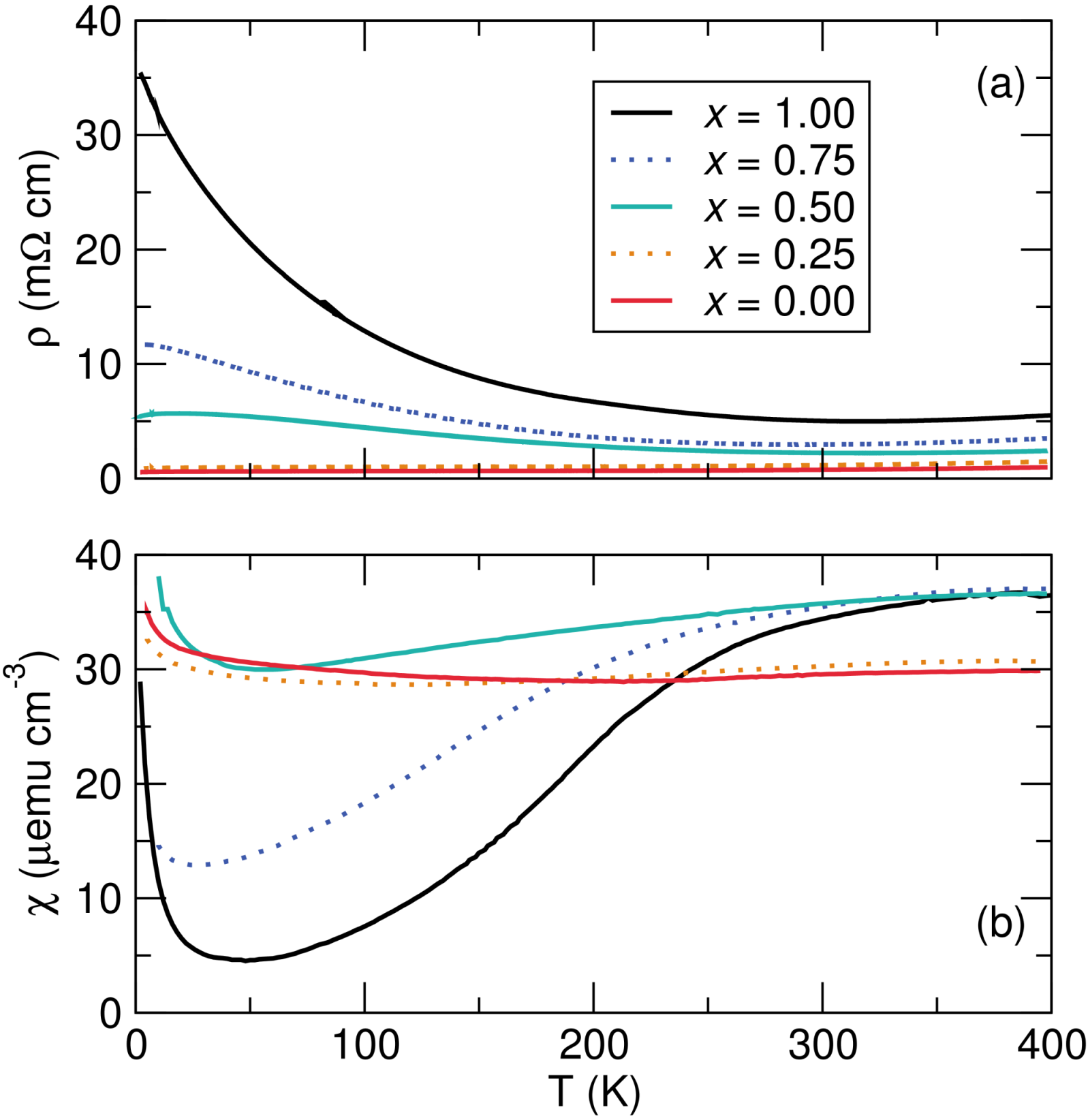}\\ 
\caption{(Color online.) Specific resistivity (a) and magnetic susceptibility (b) as 
a function of temperature for samples of 
Sr$_{1-x}$Ba$_x$Ti$_{0.9}$Nb$_{0.1}$O$_3$ 
demonstrate the metal-insulator transition as $x$ increases. The data 
in (b) have not been corrected for atomic diamagnetism.
\label{fig:rho-chi}} 
\end{figure}

The distinct properties are amply illustrated by considering the temperature 
dependence of the electrical resistivity and the magnetic susceptibility of the 
solid solution series Sr$_{1-x}$Ba$_x$Ti$_{0.9}$Nb$_{0.1}$O$_3$ displayed in 
the panels of Fig.\,\ref{fig:rho-chi}. As the value of $x$ (the amount of Ba) 
is increased across the solid solution, the resistivity 
(Fig.\,\ref{fig:rho-chi}a) displays an upturn at low temperatures; the change 
from a positive to negative temperature coefficient of resistivity takes place
between $x$ = 0.25 and $x$ = 0.50, in a region where the resistivity values   
are somewhat smaller than the value suggested by Mott for the minimum 
conductivity.\cite{Mott} In correspondence with electrical transport, the 
magnetic susceptibility of the Sr-rich side of the solid solution is 
largely temperature-independent, in keeping with its metallic transport 
properties, while the Ba-rich samples display 
a stronger temperature dependence, characteristic of moments on localized
electrons. One explanation that has been proffered for the magnetic 
behavior is the formation of bipolarons.\cite{TK_PRL}

\begin{figure}[t] 
\centering\includegraphics[width=7cm]{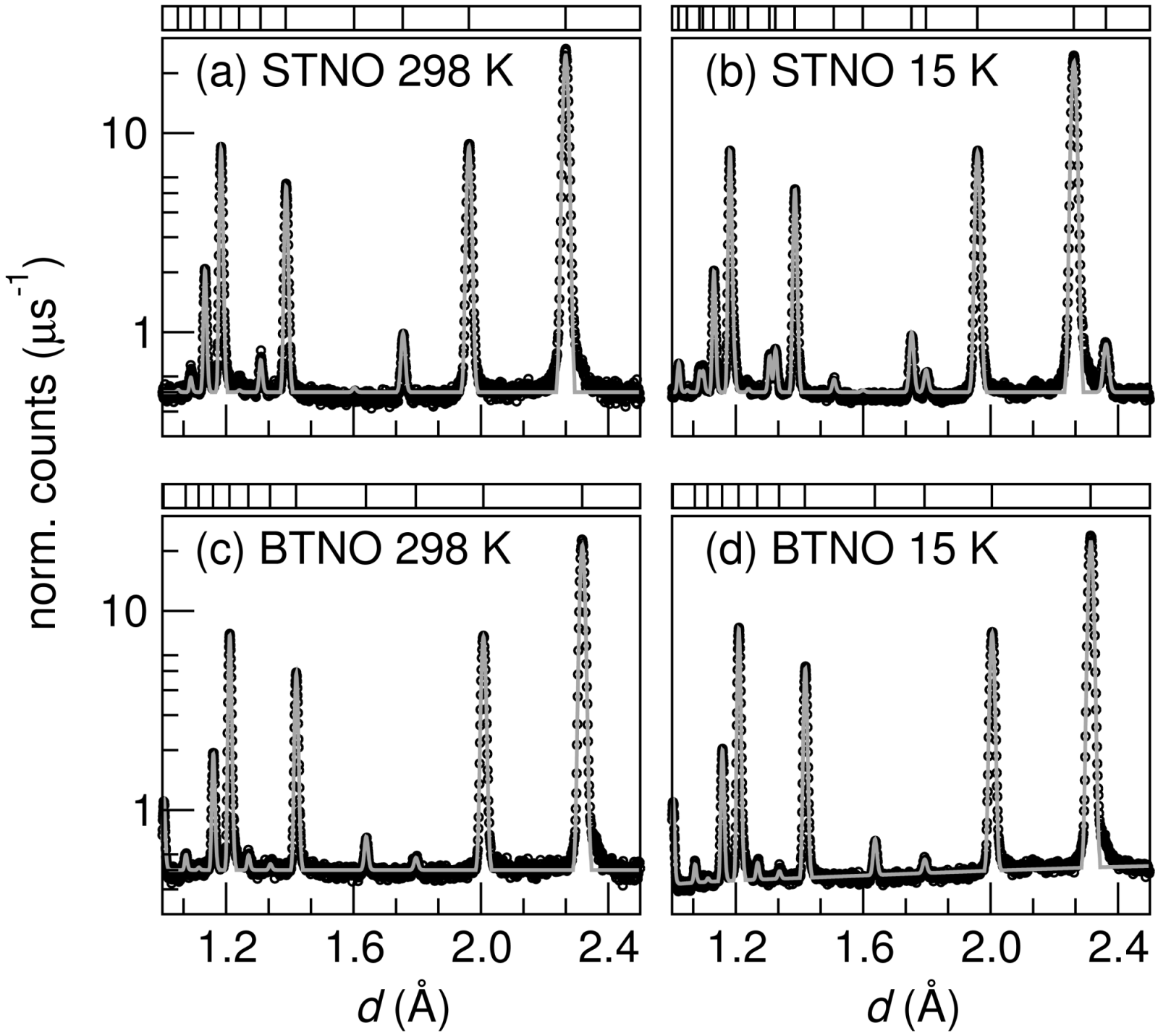}\\
\caption{A portion of the highest resolution bank of time-of-flight 
neutron diffraction data for \STNO\/ at (a) 298\,K and (b) 15\,K, and 
\BTNO\/ at (c) 298\,K and (d) 15\,K. Circles are data and the lines
are Rietveld fits to the cubic $Pm\bar3m$ structures for (a), (c), and (d),
and to the tetragonal $I4/mcm$ structure for (b). Vertical lines at the top 
of each panel indicate expected peak positions. Note the log scale.
\label{fig:rietveld}}
\end{figure}

STNO and BTNO have very similar structures and lattice parameters.  At room
temperature, both compounds crystallize in the cubic perovskite  $Pm\bar3m$
structure. Nb-substitution results in small 
expansion of the cell edge as expected from the 6-coordinate Shannon-Prewitt
radii, as Ti$^{4+}$  ($r$ = 0.61\,\AA) is either substituted by Ti$^{3+}$ ($r$ =
0.67\,\AA),  Nb$^{5+}$ ($r$ = 0.64\,\AA), or Nb$^{4+}$ ($r$ =
0.68\,\AA).\cite{Shannon} Rietveld refinement of time-of-flight neutron
diffraction data confirms the average structures of both STNO and BTNO at room
temperature are cubic. The only significant difference is  in the cell
parameters, with $a$ = 3.9237(1)\,\AA\/ and $a$ = 4.0147(1)\,\AA\/ respectively, in keeping with the much larger 12-coordinate radius of
Ba$^{2+}$ ($r$ = 1.61\,\AA) compared with Sr$^{2+}$ ($r$ = 1.44\,\AA). Rietveld
profile fits are displayed for STNO and BTNO at room temperature  in
Fig.\,\ref{fig:rietveld}a and Fig.\,\ref{fig:rietveld}c. At 15\,K, BTNO remains
cubic (Fig.\,\ref{fig:rietveld}d), while STNO transforms between 298\,K and
15\,K to the tetragonal $I4/mcm$ low temperature
structure of SrTiO$_3$ (Fig.\,\ref{fig:rietveld}b).\cite{JauchPalmer,Tucker} 
The structure is displayed in 
Fig.\,\ref{fig:structures}(b). In contrast to
SrTiO$_3$ where the octahedral rotational order parameter at
20\,K is 1.97$^{\circ}$,\cite{Tucker} the value obtained here for STNO at 15\,K
is 3.46(2)$^{\circ}$. This larger value for the Nb-substituted compound 
reflects the slightly decreased tolerance factor. 

\begin{figure}[t]
\centering\includegraphics[width=7cm]{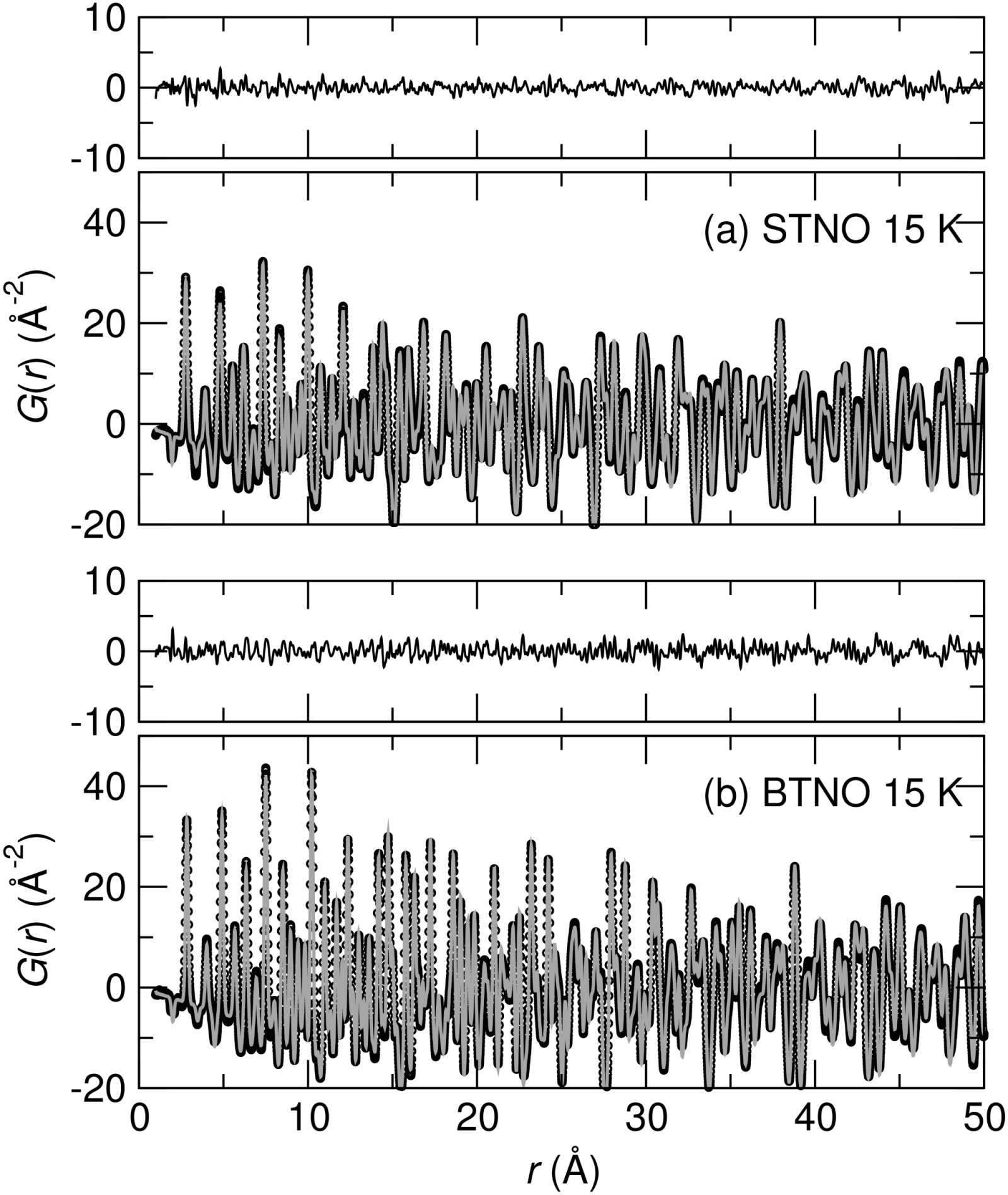}\\
\caption{ Neutron pair distribution functions (PDF) of (a) \STNO\/ 
and (b) \BTNO\/ acquired at 15\,K. Circles are data and lines are fits to the 
average tetragonal (a) and cubic (b) structures. The difference 
between data and fit is displayed separately at the top of each panel. 
\label{fig:PDF1}}
\end{figure}

\begin{table}[b]
\caption{Average structures of \STNO\/ and \BTNO\/ from 
Rietveld and 50\,\AA\/ PDF refinement of time-of-flight neutron scattering data. 
Note the anomalously high $U_{iso}$ for
the A site in the strontium compound, relieved at 15\,K \textit{via} the
phase transition to the tetragonal structure. The high $U_{iso}$ for (Ti/Nb)
for the barium compound remains at all temperatures.  Refined parameters are
shown with error.
\label{table:structure}}
\begin{tabular}{l|cc|cc}
\hline \hline
\multicolumn{5}{c}{\STNO}\\
\hline \hline 
& \multicolumn{2}{c|}{300\,K, $Pm\bar3m$} & \multicolumn{2}{c}{15\,K, $I4/mcm$}\\ 
\hline 
& Rietveld & PDF & Rietveld & PDF\\ 
\hline
$a$ (\AA)                  & 3.9237(1)  &3.9255(1)  & 5.5391(2)  & 5.5418(4) \\ 
$c$ (\AA)                  &            &           & 7.8312(4)  & 7.835(1)  \\ 
$U_{iso}$(Sr) (\AA$^2$)    & 0.0098(1)  &0.0076(2)  & 0.0053(1)  & 0.0034(1) \\ 
$U_{iso}$(Ti/Nb) (\AA$^2$) & 0.0056(3)  &0.0050(2)  & 0.0042(3)  & 0.0028(1) \\ 
$U_{iso}$(O) (\AA$^2$)     & 0.0098(1)  &0.0084(1)  & 0.0054(1)  & 0.0042(1) \\ 
$x$(O2)                    &            &  	    & 0.2349(1)  & 0.2350(1) \\ 
occ.(Nb)                   & 0.127(2)   &1/8  	    & 0.123(2)	 & 1/8       \\ 
$R_w$ (\%)                 & 3.6        &7.7 	    & 3.8 	 & 7.7       \\ 
\hline
\hline									 
\multicolumn{5}{c}{\BTNO}\\ 
\hline \hline 
& \multicolumn{2}{c|}{300\,K, $Pm\bar3m$} & \multicolumn{2}{c}{15\,K, $Pm\bar3m$}\\ 
\hline 
& Rietveld & PDF & Rietveld & PDF\\ 
\hline
$a$(\AA)                   & 4.0147(1)  & 4.0165(1)  & 4.0084(1) & 4.0100(1) \\ 
$U_{iso}$(Ba) (\AA$^2$)    & 0.0061(1)  & 0.0052(3)  & 0.0028(1) & 0.0021(1) \\ 
$U_{iso}$(Ti/Nb) (\AA$^2$) & 0.0093(3)  & 0.0080(2)  & 0.0070(2) & 0.0059(1) \\ 
$U_{iso}$(O) (\AA$^2$)     & 0.0077(1)  & 0.0061(1)  & 0.0053(1) & 0.0040(1) \\ 
occ.(Nb)                   & 0.131(1)   & 1/8	     & 0.132(1)	 & 1/8       \\
$R_w$(\%)                  & 3.0        & 7.4	     & 3.4 	 & 8.5       \\ 
\hline \hline 
\end{tabular}
\end{table}

Local structure was analyzed in terms of the atomic pair distribution function,
$G(r)$, using A$_8$Ti$_7$NbO$_{24}$ supercells, as well as single-unit cell
models with appropriate partial Ti/Nb occupancies. In Fig.\,\ref{fig:PDF1}, 
we compare the 15\,K data and PDF fits to the average tetragonal structure of 
(a) \STNO\/
and the average cubic structure of (b) \BTNO\/ to a maximum vector length of
50\,\AA. Both compounds are remarkably well described by the average structures
over this range, suggesting that local effects due to Nb-substitution are small.
A $2\times2\times2$ perovskite supercell with Nb at the center and Ti at all
corners, faces, and cell edges provided a model for probing  whether the
coordination of NbO$_6$ octahedra were distinct from those of  TiO$_6$
octahedra. Within the resolution of these experiments, no differences were
discernible, supporting the analysis using an average  \textit{B}-occupation
unit cell. Results of the PDF refinement are displayed alongside results of the
Rietveld refinement in Table\,\ref{table:structure}. Significant anomalies are
found in the isotropic displacement parameters, $U_{iso}$ for the two compounds.
In \STNO\/ it is seen at room temperature that $U_{iso}$ is somewhat large on
the \textit{A} site (Sr) compared to the  value found for Ba in \BTNO. This is a
signature of the tilting instability  that drives the cubic compound to
transform to a tetragonal ground state, whereupon all values of $U_{iso}$ at
15\,K are somewhat more  reasonable.\cite{Megaw} In \BTNO\/ on the other hand,
it is the \textit{B} 
site (Ti/Nb) that has an anomalously large $U_{iso}$, both at 300\,K and  at
15\,K.

\begin{figure}[t]
\centering\includegraphics[width=8cm]{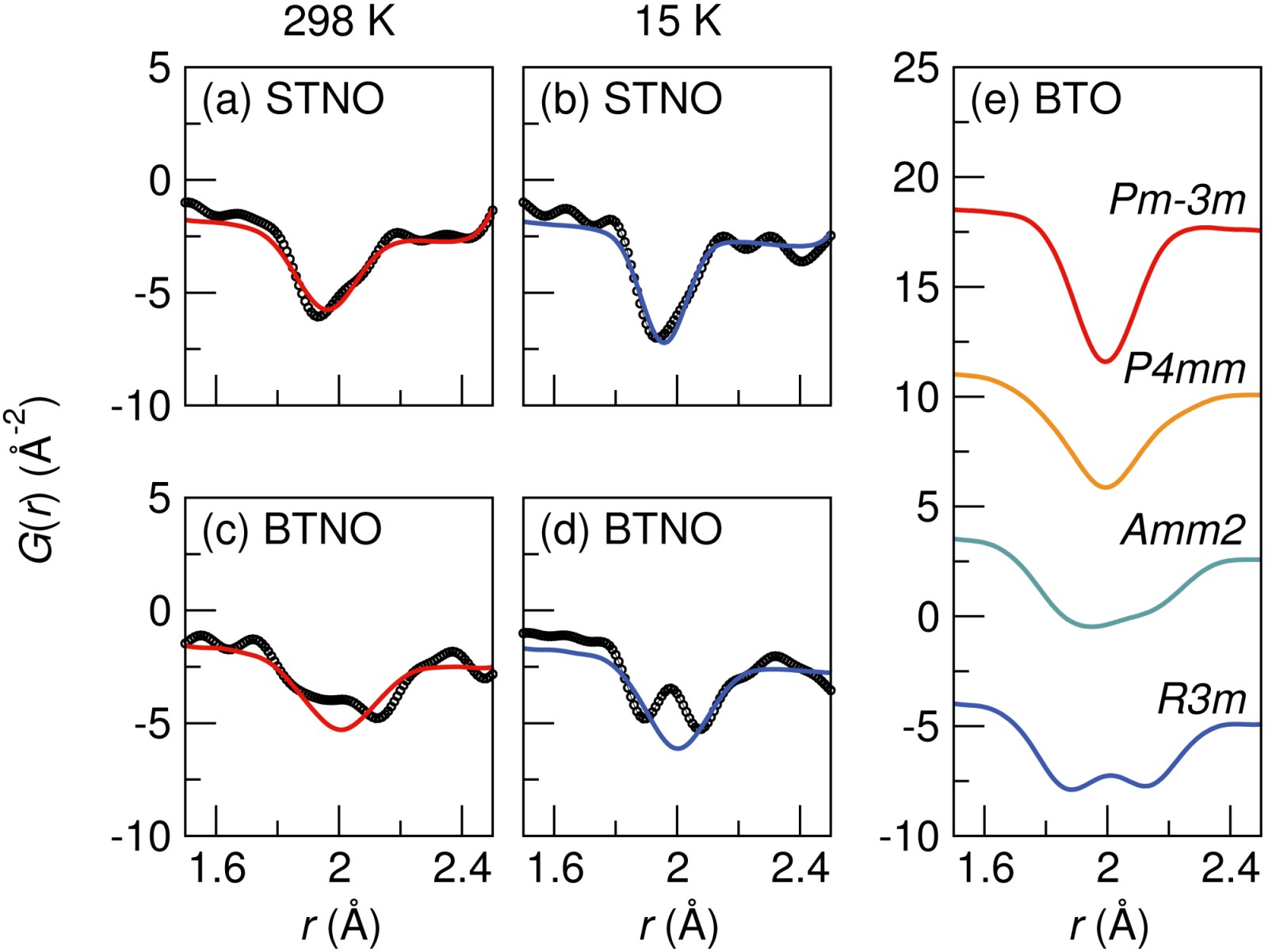}\\
\caption{(Color online.) Shortest (Ti/Nb)-O distances in the (a) 298\,K PDF and (b) 
15\,K PDF of \STNO\/. Circles are experimental data and the lines are fits to 
the respective average structures. (c) and (d) show these distances for \BTNO\/
at 298\,K and 15\,K respectively. (e) Calculated shortest Ti-O distances in
BaTiO$_3$ for the different crystallographic modifications, using the structures of 
Kwei \textit{et al.}\cite{Kwei} Note that the (Ti/Nb)-O distances are seen 
as negative peaks because of the negative neutron scattering length of Ti.
\label{fig:PDF2}}
\end{figure}

Figure\,\ref{fig:PDF2} (a through d) displays the short $r$ PDFs of \STNO\/ 
and \BTNO\/ at room temperature and at 15\,K. For \BTNO\/ at 300\,K and more markedly at 15\,K,
this first coordination shell is clearly bimodal. In contrast, the local 
structure of \STNO\/ is undistorted and well-described by the crystallographic
model. The local distortion in \BTNO\/ explains the larger \textit{B}-site 
$U_{iso}$ seen both in 
Rietveld and PDF results.  This is the principal finding of this work, and 
points to a subtle yet fundamental difference in the local structures of these 
two compounds. To better understand the bimodal first (Ti/Nb)-O peak in the 
PDF of \BTNO, we show alongside in Fig.\,\ref{fig:PDF2}(e), simulations of
the first peak PDF of BaTiO$_3$ in its different crystalline structures:
As it is cooled from high temperatures, BaTiO$_3$ transitions successively 
from cubic $Pm\bar3m$ to tetragonal $P4mm$ to orthorhombic $Amm2$ to, finally, 
it's rhombohedral $R3m$ ground state.\cite{Megaw,Kwei}
In the rhombohedral structure, Ti atoms 
in BaTiO$_3$ displace toward octahedral faces, resulting in three short and 
three long Ti-O distances, giving rise to a bimodal first peak in the 
PDF. This is precisely the situation in \BTNO. However in BaTiO$_3$ local 
distortions  are coherent and strain-coupled, and lead to phase transitions with
associated  changes in the average structure. In \BTNO, the random substitution
of Ti by  12.5-atom\% Nb is sufficient to frustrate long-range ordering, and 
the quality of \textit{cubic} refinements support the idea that displacements of
(Ti/Nb) are   very poorly correlated. 

The results of this work underpin the differences in the local 
structures of \STNO\/ and \BTNO\/ which draw from the distinct ground
states of the undoped compounds. The introduction of random potentials 
\textit{via} Nb-substitution on the Ti site in BaTiO$_3$, in addition to the
introduction of charge carriers due to the aliovalent nature of the 
substitution, result in dipole-dipole correlations being rapidly
suppressed. In particular, the role of charge carriers in bringing this
about is dramatic. In systems where the substitution is isovalent, for 
example, BaTi$_{1-x}$Sn$_x$O$_3$, ferroelectricity is suppressed only somewhat
slowly.\cite{BaTiSnO3} In keeping with ideas of ``ferroelectric metals'' 
(and their paucity), charge carriers screen the long-range electrostatic 
interactions responsible for ferroelectric 
order.\cite{anderson,sergienko} However, local effects that are responsible 
for off-centering, namely the second-order Jahn-Teller distortion on 
Ti$^{4+}$, are quite robust to changing electron counts, and persist even when 
long-range ordering of dipoles has been suppressed. The precise relation 
between local off-centering and localized electronic ground states in 
\BTNO\/ does not emerge from this study, except in that they suggest a means of 
lifting  electronic degeneracy. In \STNO, where distortions are not found,
there is no symmetry-derived mechanism for lifting the Ti/Nb $t_{2g}$ 
electronic degeneracy.

We acknowledge helpful discussions with P. A. Pincus, P. B. Littlewood, and 
Chris Van de Walle. This work has benefited from the use of NPDF at 
the Lujan Center at Los Alamos Neutron Science Center, funded by the DOE
Office of Basic Energy Sciences. Los Alamos National Laboratory is operated 
by Los Alamos National Security LLC under DOE contract DE-AC52-06NA25396.
The National Science Foundation is acknowledged for support in the
form of a Graduate Student Fellowship to KP, a Career Award to RS 
(grant DMR04-49354), and for an upgrade of the NPDF instrument at Los Alamos 
(grant DMR00-76488). TK acknowledges MEXT for financial support.


\begin{thebibliography}{99}

\bibitem{Nb_STO_Plamsons}
F. Gervais, J.-L. Servoin, A. Baratoff, J. G. Bednorz, and G. Binnig,
\textit{Phys. Rev. B} 47 (1993) 8187--8194.

\bibitem{Nb-STO_ThinFilms}
T. Tomio, H. Miki, H. Tabata, T. Kawai, and S. Kawai,
\textit{J. Appl. Phys.} 76 (1994) 5886--5900.

\bibitem{Nb_SrTiO3_supercond}
J. F. Schooley, W. R. Hosler, E. Ambler, J. H. Becker, M. L. Cohen, and 
C. S. Koonce,
\textit{Phys. Rev. Lett.} 14 (1965) 305--307.

\bibitem{Megaw}
H. D. Megaw, Ferroelectricity in crystals,  Methuen, London, 1957.

\bibitem{Marucco_solidsol} J. F. Marucco, M. Ocio, A. Forget, and D. Colson, 
\textit{J.  Alloys Compounds} 262--263 (1997) 454--458.

\bibitem{proffen_npdf}  Th. Proffen, T. Egami, S. J. L. Billinge, A. K.
Cheetham, D. Louca and J. B. Parise, 
\textit{Appl. Phys. A} 74 (2002) S163--S165.

\bibitem{gsas-expgui}  
A. C. Larson and R. B. Von Dreele, General Structure Analysis System (GSAS), 
Los Alamos National Laboratory Report LAUR 86-748 (2000);
B. H. Toby, \textit{J. Appl. Crystallogr.} 34 (2001) 210--213.

\bibitem{PDFGETN}  
P. F. Peterson, M. Gutmann, Th. Proffen and S. J. L. Billinge, PDFGETN: 
\textit{J. Appl. Crystallogr.} 33 (2000) 1192.

\bibitem{PDFgui}
C. L. Farrow, P. Juhas, J. W. Liu, D. Bryndin, E. S. Bo\v{z}in,
J. Bloch, Th. Proffen, and S. J.  L. Billinge, 
\textit{J. Phys. Condens. Matter} 19 (2007) 335219(1--7).
 
\bibitem{Mott}
N. F. Mott, Metal-insulator transitions, 2nd ed. Taylor and Francis,
London, 1990.

\bibitem{TK_PRL} 
T. Kolodiazhnyi and S. C. Wimbush, 
\textit{Phys. Rev. Lett.} 96 (2006) 246404(1--4).

\bibitem{Shannon}
R. D. Shannon, \textit{Acta Crystallogr. A} 32 (1976) 751--767.

\bibitem{JauchPalmer}
W. Jauch and A. Palmer, 
\textit{Phys. Rev. B} 60 (1999) 2961--2963.

\bibitem{Tucker} 
Q. Hui, M. G. Tucker, M. T. Dove, S. A. Wells, and D. A Keen,
\textit{J. Phys.: Condens. Matter\/} 17 (2005) S111--S124.

\bibitem{Kwei}
G. H. Kwei, A. C. Lawson, S. J. L. Billinge, and S. W. Cheong,
\textit{J. Phys. Chem.} 97 (1993) 2368--2377.

\bibitem{BaTiSnO3}
T. Nakamura and S. Nomura,
\textit{Jpn. J. Appl. Phys.} 5 (1966) 1191--1196.

\bibitem{anderson}
P. W. Anderson and E.  Blount,
\textit{Phys. Rev. Lett.} 14 (1965) 217--219.

\bibitem{sergienko}
I. A. Sergienko, V. Keppens, M. McGuire, R. Jin, J.  He, S. H. Curnoe, 
B. C. Sales, P. Blaha, D. J. Singh, K. Schwarz, and D. Mandrus,
\textit{Phys. Rev. Lett.} 92 (2004) 065501(1--4).


\end{thebibliography}
\end{document}